\documentstyle[a4,11pt]{article}

\font\tenboldmath=cmmib10

\def\beq{\begin{equation}}
\def\eeq{\end{equation}}
\def\beqry{\begin{eqnarray}}
\def\eeqry{\end{eqnarray}}

\def\si{\sigma}
\def\om{\omega}
\def\Omo{\Omega_0}
\def\vOm{{\bf\Omega}}
\def \vom {\hbox{\tenboldmath\char'041}}

\begin{document}

\title{Comments on the ERA-2005 numerical theory\\ 
of Earth rotation}
\date{\today}
\author{P.M. Mathews$^{1}$, N. Capitaine$^{2}$ and V. Dehant$^{3}$ }
\maketitle

\noindent $^{1}$ Department of Theoretical Physics,\newline
University of Madras, Chennai 600025, India \newline
e-mail: indi4272@yahoo.com \newline

\noindent $^{2}$ Observatoire de Paris, SYRTE/UMR 8630-CNRS
\newline 61, avenue de l'Observatoire, 75014 Paris, France
\newline e-mail: n.capitaine@obspm.fr \newline

\noindent $^{3}$ Royal Observatory of Belgium,
\newline 3 avenue Circulaire, B 1180 Brussels, Belgium
\newline e-mail: v.dehant@oma.be \newline

\begin{abstract}
\smallskip
Two papers recently published in Celestial Mechanics (Krasinsky 2006, and 
Krasinsky and Vasilyev 2006) have presented a model for Earth-rotation 
variations, called ERA-2005, based on numerical integration of a new set 
of equations for the rotation of a deformable Earth followed by a fit of 
the results of the integration to VLBI data. These papers claimed that 
this model was superior to any other existing model. The purpose of this 
Note is to bring to light fundamental errors in the derivation of the 
basic equations of the new theory, compounded by  serious deficiencies in 
the process of fitting to the data; they make ERA-2005 unsuitable for 
consideration as a geophysics-based model of nutation and precession. 

\end{abstract}

\section{Introduction.}

Two papers recently published in Celestial Mechanics have presented a new 
model for the variations in rotation of a deformable Earth, based on  
numerical integration of a set of rotation equations and its fit to 
VLBI data.\\

In the first of these papers, Krasinsky (2006) presented a revised  
version of the Sasao-Okubo-Saito (SOS) formulation of the equations 
governing the variations in rotation of a two-layer Earth (Sasao et al., 
1980), and used the Poincar\'e formalism to generalize it to the case 
of an Earth consisting of a mantle enclosing a fluid core with a 
hypothetical inner fluid core within it but without any solid inner core. 
Numerical integration of more general equations that take account of 
the inner fluid core layer, ocean tides, dissipative effects, etc., was 
done, after transformation to a celestial frame, for determining the 
temporal variation of the Earth rotation variables. 
Krasinsky and Vasilyev (2006) 
presented, in an accompanying paper, the results obtained by 
optimization of the fit of the numerical solutions to the time series 
constructed from a number of Very Long Baseline Interferometry (VLBI) data 
sets. (These two papers will be referred to hereunder as Papers 1 and 2.)  
The new numerical model of Earth rotation variations, thus 
constructed, goes under the name ERA-2005. 

The ERA-2005 papers express a number of criticisms of the MHB theory 
(Mathews et al. 2002, Buffett et al. 2002, Herring et al. 2002) on 
which the current IAU 2000 nutation model is based, and assert that 
ERA-2005 provides a superior and more accurate geophysical model. It is 
clearly necessary to examine the validity of these claims and criticisms. 
This is the chief motivation for this Note. 

\section{Criticisms and claims made in the ERA papers}

Krasinsky holds that the SOS equations are incorrect and need to be
revised because of errors in the original derivation, and that
these errors are reflected in the IAU 2000A nutation model since this
model is based on the MHB theory of Mathews et al. (2002) which, in
turn, is a generalization (in a number of ways) of the SOS theory. (For
brevity, we shall refer to this paper simply as MHB.) The principal 
points of the contentions in Paper 1 are (a) that the expressions employed 
by SOS for the increments to the inertia tensors of the whole Earth and 
of its core regions due to the effects of the centrifugal potentials 
associated with the Earth's variable rotation are incorrect, and (b) 
that the effects of dissipative phenomena on the responses of the Earth 
to tidal forcing are not adequately modeled in the MHB theory where they 
appear through complex increments to compliance parameters, since that 
theory fails to reveal a secular variation of the obliquity.  

The ERA papers go on to claim a smaller weighted root mean square 
(wrms) value for the residuals of the numerical output from ERA-2005 
(relative to observational data up to 2005) than the wrms of the IAU 2000
residuals relative to the same data, and even predict that the latter 
residuals in ecliptic longitude will worsen to the 2~mas level by 2009. 
On the basis of the claimed improvement in the residuals as well as the 
supposed errors in the MHB theory, they assert that the ERA-2005 model 
is the superior one.

It is also claimed that thanks to more rigorous equations, the ERA-2005 
theory makes it possible to detect the geodesic precession from the 
analysis of VLBI data, while the IAU 2000 theory would not.

We show however in the next section that the arguments for revision of 
the SOS theory are merely a reflection of fundamental misconceptions 
about the centrifugal perturbations associated with wobble motions, and 
about the deformations produced by them, and that it is Krasinsky's 
revised version of the SOS theory rather than the original SOS theory 
that is in error. We point out a couple of major consequences of this 
error, including in particular, a wrong expression for the scale factor 
relating the precession rate in longitude to the ellipticity parameter 
$e$. In succeeding sections we comment on other points of interest: 
the different ways in which dissipative phenomena are modeled in the ERA 
and MHB theories; certain strange and unphysical aspects of the Earth 
model employed; and the values  
reported for some of the parameters of ERA-2005 as estimated from fits 
to data, that are inconsistent with information from seismological and 
astronomical studies.

We also comment on problems with the claimed detection of the geodesic 
precession. 

\section{Comments on centrifugal perturbations and application of 
the Love number formalism}
 
Consider first the centrifugal potential associated with the variable 
angular velocity $\vom$ of Earth rotation (${\bf \Omega}$ of MHB) which 
consists of the constant angular velocity $\vom_0$ of rotation of the 
unperturbed Earth ($\vOm_0$ of MHB) and a very small perturbation 
$d\vom=\vom-\vom_0$ resulting from the torquing action of the tidal 
potential. It has been a well accepted concept, which is the basis of 
Clairaut's theory of the hydrostatic equilibrium Earth, that the entire 
ellipticity $e$ of the unperturbed Earth results from the incessant 
action of the time independent centrifugal potential due to the 
steady rotation $\vom_0$ 
which has nothing to do with the luni-solar perturbation. It is the small 
incremental centrifugal potential associated with the perturbation 
$d\vom$ due to the luni-solar potential (along with additional 
increments resulting from the differential rotations of the core regions 
which also result from the tidal perturbation) that produces the time 
dependent deformations and the attendant increments to the inertia tensor 
of the Earth, which react on the Earth rotation itself. 

On the other hand, it is insisted in Paper 1 that the rotational increment 
$dI_r$ to the inertia tensor should be that arising from the total 
deformation produced by the centrifugal potential associated with the 
{\it full} angular velocity $\vom$.      

Treatment of the constant part of the centrifugal potential as if it 
were part of the tidal perturbation is a grave conceptual error, which 
leaves the off-diagonal elements $c_{13}$ and $c_{23}$ that appear 
in the SOS theory unaltered but makes the diagonal elements of $dI_r$ 
many orders of magnitude larger than otherwise.  This error is 
propagated further through the use of reciprocity relations connecting 
the incremental inertia tensors $dI_r^c$ and $dI_r^i$ of the full core 
and inner core to $dI_r$ (see equation (77) of Paper 1). 

This error is compounded by a misstep in the 
application of the Love number formalism to express the increments to 
the Earth's gravitational potential that result from the above-mentioned 
deformations. 
The incremental gravitational potential at the Earth's surface is the 
appropriate $k$ Love number times the potential causing the deformation, 
which is the centrifugal potential here. The relevant Love number is 
$k_2$ in the case of time dependent degree 2 potentials; but it is the 
{\it secular Love number} $k_s$ that is appropriate when the potential 
is time independent. It has long been accepted that the behaviour 
of the Earth under a force which is sustained over a very long period is 
akin to that of a fluid body yielding to shear forces, rather than that of 
a purely elastic solid body. For example, it is mantle convection, rather 
than elastic  deformation of the mantle, that results from the stresses 
associated with persistent thermal gradients.  It is because of the 
Earth's fluid-like behavior under forcing over very long periods that the 
need arises to use the fluid/secular Love number $k_f$ or $k_s$ in going 
over from the degree 2 part of the constant centrifugal potential of 
$\vOm_0$ to the deformational contribution that it makes to the 
geopotential. The expression which relates $k_f$ or $k_s$ to the 
ellipticity $e$ of the Earth may be written down in the form
\beq
e=k_s\, \frac{M_ER^2}{3A}\,\frac{\Omo^2 R}{GM_E/R^2},
\eeq
which brings out the fact that the Earth's ellipticity $e$ is determined
by the balance between the constant centrifugal force and the 
gravitational force at the Earth's surface, as is expected in the
hydrostatic equilibrium state wherein shear resistance plays no part. It
may be noted that the need to use the fluid/secular Love number 
(with a value $\approx 0.94$)
has been pointed out in other contexts: see, for example, the third
paragraph of Section~1.1 entitled ``Treatment of the permanent tide'' in
Chapter~1 of the IERS Conventions 2003 (IERS Technical Notes 32) and more
particularly, Fig. 1.2 of the same section, concerning the contribution
from the time independent part of the degree 2 zonal tidal potential to
the geopotential.

The ERA theory fails to make the distinction between the Earth's 
short-period and long-term responses to forcing, and applies $k_2$ to 
the full potential associated with the rotation $\vom$, though only a 
tiny part of it arises from the time dependent perturbation  
part $\vom-\vom_0$ of $\vom$. As a consequence, one finds at the end of 
Sec.~3.5 of Paper 1 the strange claim that the flattening parameter $J_2$ 
of the Earth is made up of a part $\sigma J_2=(k_2/k_s)J_2\approx J_2/3$ 
which arises from the deformation caused by the centrifugal potential 
associated with the total angular velocity $\vom$, and an unforced 
part $J_2^{(0)} =(1-\sigma)J_2$ which would be present even if the 
Earth were non-rotating. (Basic physics tells us that in the absence of 
the rotation-driven potential, the balance between the gravitational 
force in the Earth and the elastic force resisting compression, without
any persistent shear resistance, would cause the shape to be spherical.) 
If the Love number $k_s$ appropriate to the effectively constant potential 
had been used, it would have been found that there was no scope at all 
for an unforced part $J_2^{(0)}$ (or for the equivalent $e_0$ referred 
to at the beginning of Sec.~4.1 of Paper 1) since $k_2/k_s$, and 
hence $\sigma$, would 
be replaced by unity; the deformation contribution to $J_2$ would then be 
all of $J_2$ (or equivalently, the ellipticity $e$ would be entirely due 
to the constant rotation), which is in accordance with the hydrostatic 
equilibrium theory that is universally acknowledged. 

The consequences of the erroneous concept concerning the centrifugal 
potential, and of the use of the wrong Love number referred to in the last 
paragraph, pervade and vitiate essentially all of the rest of the ERA 
theory and are responsible for the differences between the revised SOS 
equations developed in Paper 1 and the original equations (other than the 
dissipation terms). To give an example of the consequences: The 
luni-solar precession rate becomes proportional to the scale factor 
$e/(1+e-e\si/3)$ dependent on the deformability parameter $e\si=ek_2/k_s$ 
which is $\kappa$ in the SOS notation. (An incorrect factor  
$e/(1+e+2e\si/3)$ is shown in the numbered paragraph 4 of Sec.~1 of 
Paper~1, apparently by mistake.) However, the scale factor
which follows from all earlier treatments is $e/(1+e)$, consistent with 
the Poincar\'e concept of hydrostatic rigidity; it has remained 
unquestioned so far.  (Though the Poincar\'e treatment of nutation 
employed an Earth model made up of a rigid shell enclosing a uniform  
fluid core, the very same scale factor emerges transparently from 
the SOS or MHB equations which take full account of the Earth's 
deformability as well as the stratification of the fluid core. It is 
stated mistakenly in the above-cited paragraph of Paper~1 that the SOS 
theory would lead to the scale factor $e/(1+e+e\si)$ in the case of a 
deformable Earth.) The dependence of the ERA scale factor on  
deformability is simply a consequence of the conceptual errors in the 
revised SOS equations that have been pointed out above. 

Another consequence of the revision is the new expression for the 
frequency of the Free Core Nutation (FCN) mode given in Appendix A.1 of 
Paper 1. When translated to the SOS notation, it is $(A/A_m)\Omo 
(e_c-\beta-\gamma/3)$. Ignoring the choice of the positive sign for 
$f_{FCN}$ as noted below equation (150), which makes no sense, we observe 
that the presence of the new term $\gamma/3$ with a value $\approx e_c/4$ 
represents a significant deviation from the 
original SOS formula. The FCN frequency is critical in determining 
the magnitude of the nutations, and the results of past estimates of its 
value from fits to data have been quite robust. So it is very surprising 
that it has had to be taken as an external input into the ERA fitting 
process. As stated in the paragraph numbered 2. in Appendix A.1 of Paper 
1, the 431 day period that ``is reliably estimated from the analysis of VLBI 
observations'' in the earlier literature was used as a ``a strong 
constraint'', because the value calculated from the ERA parameter 
estimates turned out otherwise to be 415 days as stated in the last 
paragraph of Sec.~3.3 of Paper 2--and this is unacceptably far from the 
reliable estimates. 

\section{Comments on the modeling of the dissipative phenomena}

In regard to the effects of dissipative phenomena, they are lumped, in 
Paper 1, into a time delay $\tau$ between the action of the tidal 
potential and the deformational response of the Earth as a whole, which is  
converted into an equivalent phase delay $\delta=\om\tau$. (Different 
phase delays $\delta_c,\delta_i$ are assigned to the core regions.) The 
inspiration for this 
comes from the explanation for the observed slow secular decrease of the 
Earth's axial rotation rate and the slow increase of the Moon-Earth 
distance, both ascribed to a continuous transfer of the Earth's rotational  
energy into the energy of orbital motion of the Moon because of the time 
delay between the tidal forcing by the Moon and the deformational 
response. 

In the MHB paper, on the other hand, specific phenomena 
which cause the responses to tidal forcing to become out of phase are 
individually modeled, based on concrete information available from 
observations: from seismic normal modes and propagation of seismic waves,  
for the modeling of anelasticity effects; and from space geodetic 
observations of the  perturbations of geopotential coefficients, for the 
incremental inertia tensor elements due to the ocean tides. It is true 
that MHB did not consider the possibility of any contribution to the 
obliquity rate. The intent of their paper was only to construct a
theoretical model for the precession in longitude and the nutations; 
comparison of the theoretical predictions with observations was also 
for these quantities only,
though the obliquity rate correction to the IAU 1976 model was estimated
empirically by Herring et al. (2002) from VLBI observations along with the 
other quantities. The computation of the second order contributions to 
nutation and
precession from the torques produced by the action of the tidal potential
on the deformations produced by this potential (in Appendix A of MHB)
was incomplete. Lambert and Mathews (2006) made a complete calculation 
of these effects (including the effect on the obliquity rate) through a 
treatment which automatically takes account of the phase shifts arising 
from anelasticity, ocean tide effects and boundary couplings of 
the core. Their result for the tidal contribution to the obliquity rate 
(0.00127 mas/yr) is much smaller than the 0.024 mas/yr found by Williams 
(1994) by a method based on the transfer of angular momentum between the 
rotational motion of the Earth and the orbital motion of the Moon keeping 
the total of the two constant. The sources of this discrepancy have 
not yet been identified.

Though we have shown above that the major criticisms of the MHB 
theory made in the ERA papers are based on misconceptions which vitiate 
the basic equations of the ERA-2005 theory, it seems desirable still to 
examine other relevant aspects of the physical modeling and the 
optimization of the fit of the output from the theory to the VLBI 
nutation-precession time series, especially in view of the claim made 
that the wrms of residuals of its fit to  VLBI-based series is lower than 
that of IAU 2000, and hence that ERA-2005 provides a superior model for 
nutation and precession. 

\section{Comments on the Earth model: Inner fluid core; Ocean tides}

The Earth model that is employed in ERA has a fluid core containing an  
inner fluid core within, while the solid inner core, with dimensions and 
properties that are pretty well determined from seismological 
observations, is neglected.  The so-called fluid inner core is totally 
nebulous in nature. With no basis in any Earth model, the ``preliminary'' 
values of its parameters as shown in Sec. 2.1 are arbitrary choices. The 
ratio of its moment of inertia to that of the whole Earth, as well as 
its ellipticity, seem to be chosen to be just about the same as those of 
the conventional solid inner core. Its existence is postulated for the 
sole purpose of providing an explanation for a free wobble mode with a 
period of about 420 days (in addition to the well known FCN mode) that is 
claimed to have been found in the residuals of the VLBI nutation series 
relative to IAU 2000. The proposed new mode (which has been named as FICN 
though it has nothing to do with the FICN mode in the earlier literature 
which is associated with the solid inner core) is supposed to explain 
the apparent ``beats'' with the 431-day FCN mode in the residuals (Fig.~3 
of Paper 2). But the beat frequency, which is half the difference between 
the individual mode frequencies, would correspond to a period of over 90 
years. It would take observations over at least half that period, 
theoretically, to separate the two modes. Even then, the amplitude of the 
free FICN would have to be of comparable magnitude to that of the FCN, 
and both the amplitudes would have to remain essentially constant over a 
long time span of the above order if the beats are to be recognizable as 
such--and neither of these conditions is physically reasonable, 
since the inner core is visualized as being very much smaller than the 
whole core, and the geophysical mechanisms available for the excitation 
of the free modes are highly variable in time. In these circumstances,  
the postulated fluid inner 
core merely serves as a device for the introduction of a couple of 
adjustable parameters ($T_{FICN}$ and $\nu_{uv}$ in Table~3 of Paper 2) 
into the fitting process. 

Another aspect of the modeling that is bizarre and physically  
unacceptable is the representation of ``the combined action of oceanic 
tides and non-uniform structure of the Earth's interior'' by an increment 
$\delta k_2=k_2^{(1)} \cos\theta_B +k_2^{(2)}\cos^2\theta_B$ to the Love 
number parameter $k_2$ (Sec.~3.3 of Paper 1), where $(\pi/2-\theta_B)$ is 
the geographical latitude of the position of the perturbing celestial 
body $B$ (Moon/Sun). It is incomprehensible how the perturbation of 
the Earth's gravitational potential by ocean tidal and other mass  
distributions which have a complicated spatial variation could be 
represented by a spatially uniform increment $\delta k_2$, and how 
the incremental potential {\it everywhere} outside the Earth could be 
determined by a scale factor which has one value, $k_2+k_2^{(1)}\sin\beta 
+k_2^{(2)}\sin^2\beta$, when the body is over a northern latitude 
$\beta=\pi/2-\theta_B$ and a different value $k_2-k_2^{(1)}\sin\beta 
+k_2^{(2)}\sin^2\beta$, when the body is over the corresponding southern 
latitude $-\beta$, with no reference to the nature of the non-uniform 
mass distributions. With no physical justification, the only role 
that the constant parameters $k_2^{(1)}$ and $k_2^{(2)}$ seem to have is 
to serve as convenient additional adjustable parameters for the fitting. 

It may be remarked that a superficially similar expression for the effect 
of the ocean tides is found in Kaula (1969). However, the dependence 
of $k_2$ in that work is not on the angular position of the celestial 
body $B$ in the terrestrial frame, but on that of the location ${\bf r}$ 
at which the incremental geopotential is to be evaluated. The latter 
dependence is unobjectionable: even for an oceanless Earth, the latitude 
dependence of the Love numbers on account of the equatorial bulge is 
well known.    

\section{Comments on the estimated parameters}
The fitting of the output from the ERA-2005 theory to the 
precession-nutation time series constructed from VLBI observations is 
done by varying the values of a large number of parameters that are 
treated as adjustable; they include 14 geophysical parameters (among  
which the {\it ad hoc} parameters $k^{(1)}$ and $k^{(2)}$ appear), 6 
empirical parameters (with admittedly unclear physical significance),  
and the initial values of the three Earth orientation parameters in 
space. Among the empirical parameters, four are adjustments to the 
4 coefficients pertaining to the retrograde and prograde 
annual nutations (see Table~4 of Paper 2). (Adjustments to just the 2 
prograde annual nutation coefficients were made in MHB. Other empirical 
parameters used were in the modeling of the frequency dependence of 
ocean tide admittances; no free parameters were used in the theoretical 
computation of the effect of the ocean tidal components on nutations. One 
other parameter that was introduced was for minor fine-tuning of 
the anelasticity model.)  Returning to the other empirical parameters in 
the ERA theory, we note that $E_1$ is intended to ``correct'' the ratio 
of the amplitudes of the 
retrograde and prograde 18.6 year nutations, and $E_2$ to ``correct'' the 
out of phase nutations. These are {\it ad hoc} corrections to bring down 
the wrms of residuals below what the geophysical parameters alone could 
accomplish. When comparing the plots of the residuals calculated with 
ERA-2005 and IAU 2000, the authors recognize (first paragraph of Sec.~3.3 
of Paper 2) that accounting for the free oscillations in ERA-2005 is the 
main cause why the wrms errors for ERA-2005 are less than those for IAU 
2000, the free nutations being not included in the latter theory. 

One finds among the estimated geophysical parameters both $e$ and 
$k_2=k_s\si$. Since the ERA scale factor for the precession rate is a 
function of $e$ and $\si$, it is evident that estimates of these two 
parameters must be highly correlated.  Again, both $e_c$ and  
$\kappa_{el}$ are 
among the estimated parameters, where $\kappa_{el}$ is the counterpart of 
${\rm Re}\,  K^{CMB}$ of MHB. Since the inertial coupling with $e_c$ as 
coefficient and the CMB coupling represented by $\kappa_{el}$ are both 
proportional to the differential wobble between the mantle and the core, 
these two parameters should be 100\% correlated, and it is not at all 
clear as to how they can both be estimated parameters. In the MHB work, 
the combination $(e_f+{\rm Re}\, K^{CMB})$ and ${\rm Im}\, K^{CMB}$ were 
among the estimated parameters. A separate estimate for ${\rm Re}\, 
K^{CMB}$ could however be obtained, thanks to the relation between it and 
${\rm Im}\, K^{CMB}$ that the theory of the electromagnetic coupling at 
the CMB called for; and the estimate for $e_f$ followed then from the 
estimate for the sum. 

A list of estimated values obtained from the fit for 14 parameters  
is found in Table~3 of Paper 2. The authors do not make any effort to 
justify these values in the light of what is known otherwise about the 
main parameters; they say simply: ``In the present work, we do not discuss 
physical meaning of the geophysical parameters $\cdots$''. Such a 
discussion is essential, however, to assess  how geophysical the theory is. 
The following remarks are offered in that spirit.

1. The estimate shown for $e$ in the Table is $3.283410\times 10^{-3}$, 
which is not consistent with any other estimates from astronomical 
observations. It is lower, by 0.0346\%, than the MHB estimate $e=  
3.284548\times 10^{-3}$. Knowing also that the first order lunisolar 
precession rate of MHB is about $5040.7''$/cy, one can easily infer 
the corresponding ERA value by multiplying the above rate by the ratio 
of the ERA scale factor $e/(1+e-e\si/3)$ (evaluated using the values from 
Paper~2 for $e$ and $\si$) to the MHB scale factor $H_d$ taken with the 
MHB value for $e$. This ratio turns out to be less than unity by $2.84\times 
10^{-5}$. Consequently the ERA precession rate must be less than the MHB 
value by about $0.14''$/cy. This difference should have been easily 
discernible as a rather steep slope in the difference (ERA-2005 $-$ IAU 
2000) in the top curve in Fig.~3 of Paper 2; why it does not, remains 
incomprehensible. 

Unfortunately, the ERA-2005 estimate for the precession rate is not given 
explicitly anywhere. The available information about it is from the 
penultimate sentence of the first paragraph of Sec.~3.3 of Paper 2 which 
says that the difference between ERA 2005 and IAU 2000 in the secular 
trend in the longitude variable (i.e., in the precession rate in  
longitude) is $(-0.82\pm 0.22)$~mas/cy. This very small value is 
grossly inconsistent with the difference of about $-0.14''$/cy $=-140$ 
mas/cy mentioned above. 

2. In regard to other parameters: it is observed that the ERA-2005  
estimate for $\alpha$ (which stands for $A_c/A$) is about 4\% less than 
the value computed from recent Earth models like PREM. To our 
understanding, seismological Earth models do not provide so much of a 
leeway in the value of this ratio. The ERA estimate for $k_2$ is quite 
unrealistic, being about 10\% less than the accepted value of about 0.3 
that recent Earth models lead to on integration of their deformation 
equations; and the estimated value of $e_c$ ($e_c=3.3761 \times 
10^{-3}$), which is seen to be even higher than that of $e$, is 
definitely unphysical: the hydrostatic equilibrium structure requires 
$e_c$ to be only about three-fourths of $e$, and there is no conceivable 
mechanism that could bring about such a gross deviation from this value as 
the ERA fit requires. The authors of the ERA papers seem to be quite 
disdainful of seismologically constructed Earth models and to consider 
themselves unconstrained by parameter values based on such models. 

All in all, it should be abundantly clear that the ERA-2005 
nutation-precession model cannot be considered to be based on sound  
geophysics, and that there is a strong internal inconsistency (noted in 
paragraph 1. above) between the precession rate dictated by the estimated 
value of $e$ taken together with the scale factor, and the claim of a 
very small wrms of the residuals for ERA-2005. 

\section{Comments on astronomical quantities used or estimated}

Besides the problems mentioned in the last two sections in regard 
to the geophysical and empirical parameters estimated and the estimates 
obtained, there are also problems with some of the astronomical 
quantities involved and with the estimates obtained in ERA 2005 for 
some of them. 

1. It is stated in Paper 1 (paragraph 4. of Sec.~1) that the ERA-2005 
theory has made possible, for the first time, a direct determination of 
the geodesic precession from the analysis of VLBI data, due to the 
superiority of the equations on which the model is based over other 
theories.

The claim of superiority of the ERA theory has been shown in earlier 
sections to be not valid. 

It may be thought possible, in principle, to estimate $H_d$ (or $e$) and 
the geodesic precession $p_g$ simultaneously from observational data on 
nutation and precession since the observed precession rate involves both the 
predominant part due to $H_d$ (about 5040.7''/cy) and the relatively small 
$p_g$ (about $-1.92''$/cy), while nutation amplitudes depend on $H_d$ 
and are independent of $p_g$. But, in reality, due to a number of practical 
reasons, separating the geodesic precession from the other contributions 
to the precession rate from observations alone is very difficult. That is why 
precession-nutation models hitherto have used theoretical values for $p_g$, 
while $H_d$ (or $e$) has been estimated from precession-nutation 
observations.  
 
Nevertheless, Paper~2 (Sec.~3.4) claims to have actually confirmed the 
effect of the geodesic precession from a fit of the ERA-2005 theory to 
VLBI data. The validity of that claim is seriously questionable for the 
following reasons.
\begin{itemize}
\item[-] The discrepancy (of about $-$0.15''/cy, i.e.~$\approx$ 10\%) 
 			of the estimated value for $p_g$ with respect to the theoretical 
 			value is reported (Sec.~3.4) to be compensated by corrections to 
 			other parameters (mainly to $e$), which clearly indicates a 
 			significant correlation between $p_g$ and other estimated parameters.  
 			Unfortunately, nothing is said in the ERA papers about the 
		 degree of correlation or about the magnitude of the changes in the 
 			estimated geophysical parameters when $p_g$ is estimated.  

\item[-] The fact that the estimate shown for $e$ in Table~2 of 
        Paper~2 is not consistent with any other estimates from astronomical  
        observations makes doubtful the possibility of a reliable 
        estimation of both $e$ and $p_g$.                
\end{itemize}

2. Another problem is related to the celestial reference system to which 
the Euler angles considered in the ERA model are referred. 
Paper~2 actually leaves it unclear whether the reported values for the Euler angles 
(e.g. Table~2) are referred to the inertial mean ecliptic frame at J2000.0 
(e.g.~Sec.~2, paragraphs~1 and 2, or Sec.~3.2), or to the geocentric ecliptic 
reference system derived from the GCRS (Geocentric Celestial Reference System) 
through a rotation $\theta_0$ for the J2000 mean obliquity around the 
x-axis (cf.~Sec.~2, paragraph 1). Moreover, Paper~2 does not make clear 
(eg.~Sec.~3.1, 2nd paragraph) how the obliquity value $\theta_0$ is 
determined.        

\section{Concluding remarks}
The assertion in the ERA-2005 papers about higher accuracy is based 
solely on a claim (a) that numerical integration of the set of equations 
of the ERA theory leads to lower residuals relative to the VLBI-based 
observational data than the residuals of IAU 2000A which is based on the 
MHB theory, and (b) that especially for the last couple of years of the 
data set used, the predictions from the MHB theory for the  
nutation-precession in longitude, in particular, show increasing 
deviations from the data. No other information that would have been of 
value in making comparisons between ERA and other theories has been made 
available. The ERA estimates for the precession rate in longitude or for 
the coefficients of any of the spectral components of nutation, which are 
quantities of the greatest interest, are not shown. No information is 
given on the magnitudes of the contributions to nutation from ocean tides 
or from the core mantle boundary coupling. Nothing has been said about 
the amplitudes and phases of the FCN or FICN oscillations, whether they 
are constant or variable, or whether excitation of these free modes by 
geophysical mechanisms are allowed for, though these oscillations are 
stated to be included in the ERA model unlike in MHB and other theories.  
And finally, there is no discussion or attempt at justification of the 
values of the geophysical parameters involved in the theory (estimated 
from fits to data or assumed a priori), most of which differ to an 
unreasonably large extent from numbers computed from seismological Earth 
models or estimated from earlier determinations. 

To sum up, the 
comparisons of the results of the ERA theory with those of other theories 
consist of nothing more than graphs, as functions of time, of the 
residuals of the Euler angles computed from the ERA theory relative to 
VLBI data and to the IAU model; and, of course, the rms of the residuals. 
We have brought to light, in this Note, fundamental errors in the derivation
of the basic equations of the ERA-2005 theory and unphysical aspects of
the geophysical modeling, compounded by serious deficiencies in the process
of fitting to the data and in the estimates obtained for 
various parameters. It should be clear therefore that ERA-2005 does not 
constitute a sound geophysics-based model of nutation and precession.

\section*{References}
 
Buffett B. A., Mathews, P. M., Herring, T. A., 2002, ``Modeling of nutation 
and precession: Effects of electromagnetic coupling'', J. Geophys. Res. 107, 
B4, 10.1029/2000JB000056

\vskip 5mm
\noindent Herring, T. A., Mathews, P. M., Buffett B.A., 2002, ``Modeling of nutation 
and precession: Very long baseline interferometry results'', J.
Geophys. Res. 107, B4, 10.1029/2001JB000165

\vskip 5mm

\noindent Kaula, W. M., 1969, ``Tidal friction with latitude-dependent 
amplitudes and phase angle'', Astron. J., 74, 1108-1114

\vskip 5mm

\noindent Krasinsky, G. A., 2006, ``Numerical theory of rotation of the 
deformable Earth with the two-layer fluid core. Part 1: Mathematical 
model'', Celes. Mech. and Dyn. Astr. 96, 3-4, 169-217

\vskip 5mm

\noindent Krasinsky, G. A., Vasilyev, M. V., 2006, ``Numerical theory of 
rotation of the deformable Earth with the two-layer fluid core. Part 2: 
Fitting to VLBI data'', Celes. Mech. and Dyn. Astr. 96, 3-4, 219-237

\vskip 5mm 

\noindent Lambert, S. B., Mathews, P. M., 2006, ``Second-order torque on 
the tidal redistribution and the Earth's rotation'', A\&A 453, 363

\vskip 5mm

\noindent Mathews, P.M., Herring, T.A., Buffett B.A., 2002,
``Modeling of nutation and precession: New nutation series for
nonrigid Earth and insights into the Earth's interior'', J.
Geophys. Res. 107, B4, 10.1029/2001JB000390

\vskip 5mm

\noindent IERS Conventions 2003, IERS Technical Note 32,
D.D.~McCarthy and G.~Petit (eds), Frankfurt am Main: Verlag des
Bundesamts f\"ur Kartographie und Geod\"asie, 2004

\vskip 5mm

\noindent Sasao, T., Okubo, S., Saito, M., ``A simple theory on the 
dynamical effects of a stratified fluid core upon nutational motion of the 
Earth'', in Proceedings of IAU Symposium 78, E. P. Federov, M. L. Smith, 
and P. L. Bender (eds), pp. 165-183, Hingham, Mass., D. Reidel, 1980     

\end{document}